\documentclass{appolb}
\usepackage{epsfig}
\usepackage{graphicx}
\usepackage{amsfonts}
\usepackage{amssymb}
\usepackage{amsmath}

\usepackage{ulem}
\usepackage{color}

\renewcommand\sout{\bgroup \color{blue} \ULdepth=-.5ex \ULset}

\def\Slash#1{\not\!\!#1}

\begin{document}
\title{Interplay between Deconfinement and Chiral Properties
\thanks{Presented at the international workshop on Critical Point and Onset of Deconfinement (CPOD 2016), May 30 - June 4, 2016, Wroclaw, Poland}%
}
\author{Hideo Suganuma, Takahiro M. Doi, 
\address{Department of Physics, Kyoto University,
Kyoto 606-8502, Japan}
\and
Krzysztof Redlich, Chihiro Sasaki, 
\address{
Institute of Theoretical Physics, University of Wroclaw, 
PL-50204 Wroclaw, Poland
}
}
\maketitle
\begin{abstract}
We study interplay between confinement/deconfinement and chiral properties.
We derive some analytical relations of the Dirac modes with 
the confinement quantities, such as 
the Polyakov loop, its susceptibility and the string tension.
For the confinement quantities, 
the low-lying Dirac eigenmodes are found to give 
negligible contribution, 
while they are essential for chiral symmetry breaking.
This indicates no direct, one-to-one correspondence 
between confinement/deconfinement 
and chiral properties in QCD.
We also investigate the Polyakov loop in terms of the eigenmodes of 
the Wilson, the clover and the domain-wall fermion kernels, respectively.

\end{abstract}

\def\slash#1{\not\!#1}
\def\slashb#1{\not\!\!#1}
\def\slashbb#1{\not\!\!\!#1}

\section{Introduction}

The relation between quark confinement and 
spontaneous chiral-symmetry breaking has been 
a longstanding difficult problem remaining in QCD physics. 
In this paper, considering the essential role of 
low-lying Dirac modes to chiral symmetry breaking \cite{R12}, 
we derive analytical relations between the Dirac modes 
and the confinement quantities, e.g., the Polyakov loop \cite{DSI14}, 
its fluctuations \cite{DRSS15} and the string tension \cite{SDI16}.
We mainly use the lattice unit, $a=1$.

\section{Dirac operator, Dirac eigenvalues and Dirac modes}

We use an ordinary square lattice with spacing $a$ and 
size $V \equiv N_s^3 \times N_t$, and impose 
the temporal periodicity/antiperiodicity for gluons/quarks.
In lattice QCD, the gauge variable is expressed as 
the link-variable $U_\mu(s)$=${\rm e}^{iagA_\mu(s)}$, 
and the simple Dirac operator is given as 
\begin{eqnarray}
\Slash{\hat D}
=\frac{1}{2a}\sum_{\mu=1}^{4} \gamma_\mu (\hat U_\mu-\hat U_{-\mu}),
\label{eq1}
\end{eqnarray}
where the link-variable operator 
$\hat U_{\pm \mu}$ is defined by \cite{DSI14,DRSS15,SDI16}
\begin{eqnarray}
\langle s |\hat U_{\pm \mu}|s' \rangle 
=U_{\pm \mu}(s)\delta_{s\pm \hat \mu,s'}, 
\end{eqnarray}
with $U_{-\mu}(s)\equiv U^\dagger_\mu(s-\hat \mu)$.
For the anti-hermitian Dirac operator $\hat{\Slash D}$ satisfying 
$\hat{\Slash{D}}_{s',s}^\dagger$=$-\hat{\Slash{D}}_{s,s'}$,
we define the Dirac mode $|n \rangle$ 
and the Dirac eigenvalue $\lambda_n$, 
\begin{eqnarray}
\hat{\Slash{D}} |n\rangle =i\lambda_n |n \rangle \quad (\lambda_n \in {\bf R}), \qquad
\langle m|n\rangle=\delta_{mn}, \qquad \sum_n |n \rangle \langle n|=1.
\end{eqnarray}

\section{Polyakov loop and Dirac modes on odd-number lattice}

We here use a temporally odd-number lattice \cite{DSI14,DRSS15,SDI16}, 
where the temporal lattice size $N_t(< N_s)$ is odd. 
In general, only gauge-invariant quantities 
such as closed loops and the Polyakov loop 
survive in QCD, according to the Elitzur theorem \cite{R12}.
All the non-closed lines are gauge-variant 
and their expectation values are zero.
Now, we consider the functional trace 
\cite{DSI14,SDI16}, 
\begin{eqnarray}
I \equiv {\rm Tr}_{c,\gamma}(\hat U_4\hat{\Slash{D}}^{N_t-1})
=\sum_n\langle n|\hat{U}_4\Slash{\hat{D}}^{N_t-1}|n\rangle
=i^{N_t-1}\sum_n\lambda_n^{N_t-1}\langle n|\hat{U}_4| n \rangle, 
\end{eqnarray}
where 
${\rm Tr}_{c,\gamma}\equiv \sum_s {\rm tr}_c 
{\rm tr}_\gamma$, 
and we use the completeness of the Dirac mode. 

From Eq.(\ref{eq1}), 
$\hat U_4\hat{\Slash{D}}^{N_t-1}$ 
is expressed as a sum of products of $N_t$ link-variable operators.
Then, $\hat U_4\hat{\Slash{D}}^{N_t-1}$ 
includes many trajectories with the total length $N_t$, 
as shown in Fig.~1.
Note that all the trajectories with the odd-number length $N_t$ 
cannot form a closed loop 
on the square lattice, and give gauge-variant contribution, 
except for the Polyakov loop.
Thus, in 
$\langle I \rangle
=\langle {\rm Tr}_{c,\gamma}(\hat U_4\hat{\Slash{D}}^{N_t-1}) \rangle$,
only the Polyakov-loop can survive 
as the gauge-invariant component, and 
$\langle I \rangle$ is proportional to the Polyakov loop $\langle L_P \rangle$.
Actually, we can mathematically derive 
the following relation \cite{DSI14,SDI16}:
\begin{eqnarray}
\langle I \rangle
&=& \langle {\rm Tr}_{c,\gamma} (\hat U_4 \hat{\Slash D}^{N_t-1}) \rangle
=\langle {\rm Tr}_{c,\gamma} \{\hat U_4 (\gamma_4 \hat D_4)^{N_t-1}\} \rangle
=4 \langle {\rm Tr}_{c} (\hat U_4 \hat D_4^{N_t-1}) \rangle 
\nonumber \\
&=&
\frac{4}{2^{N_t-1}} \langle {\rm Tr}_{c} \{ \hat U_4^{N_t} \} \rangle
=-\frac{4N_cV}{2^{N_t-1}} \langle L_P \rangle,
\end{eqnarray}
where the last minus reflects the temporal antiperiodicity of 
$\hat{\Slash{D}}$ \cite{SDI16}.

\begin{figure}[h]
\begin{center}
\hspace{-0.4cm}
\includegraphics[scale=0.435]{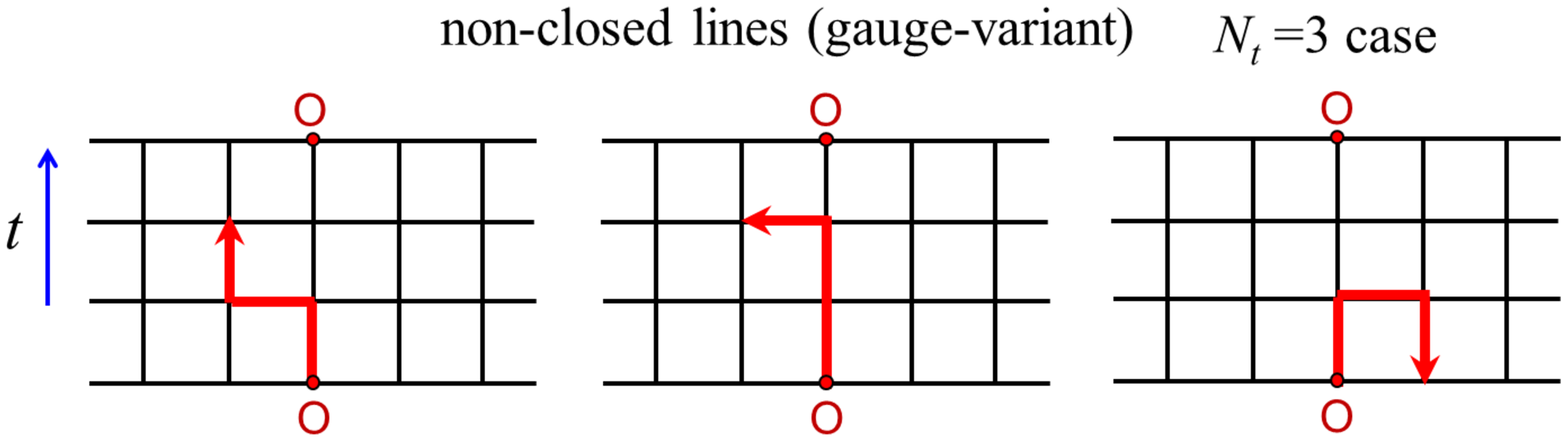}
\includegraphics[scale=0.275]{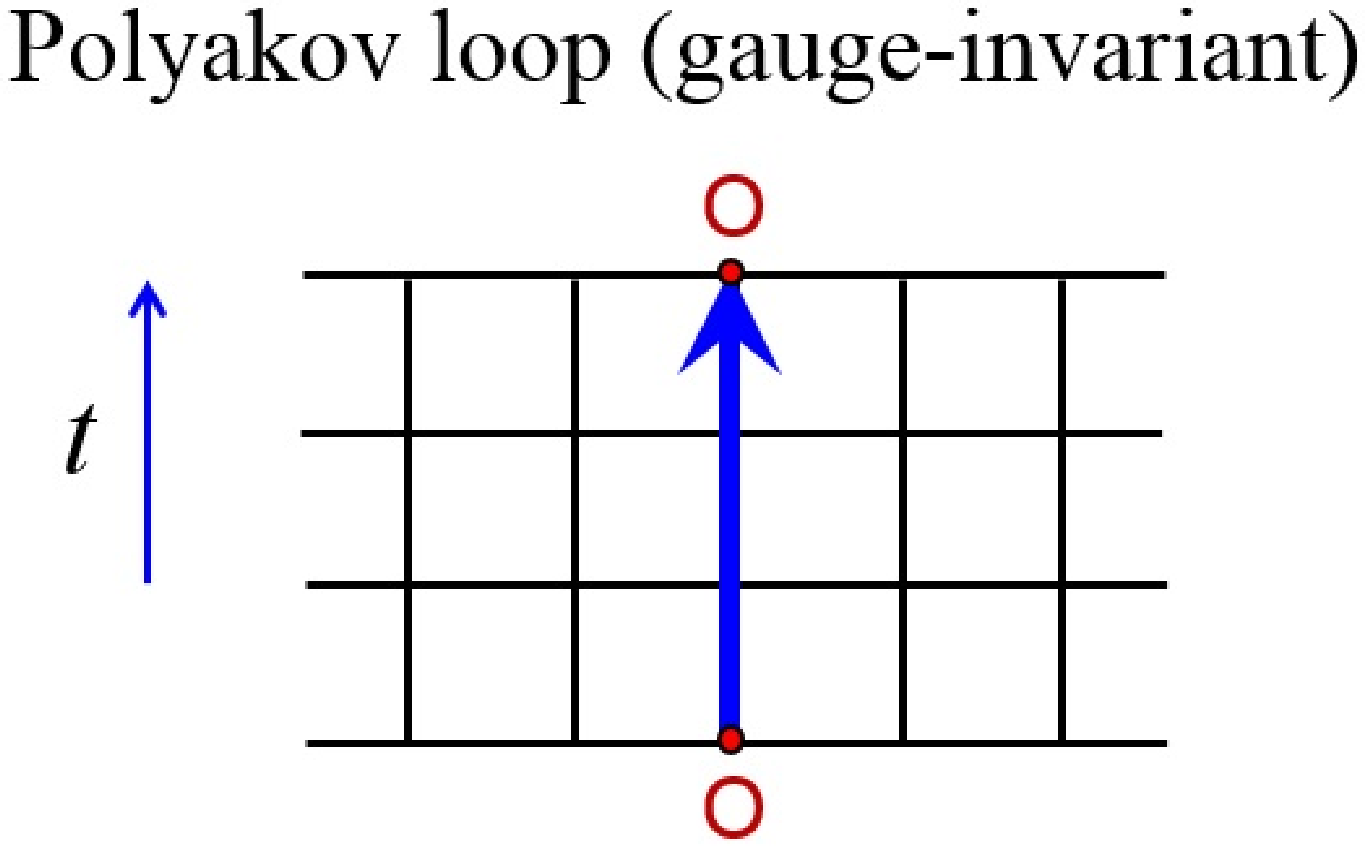}
\caption{
Examples of the trajectories stemming from 
$I ={\rm Tr}_{c,\gamma}(\hat U_4\hat{\Slash{D}}^{N_t-1})$. 
For each trajectory, the total length is 
$N_t$, and the ``first step'' is positive 
temporal direction, $\hat U_4$.
All the trajectories with the odd length $N_t$ 
cannot form a closed loop on the square lattice,
so that they are gauge-variant and give no contribution, 
except for the Polyakov loop.
Thus, only the Polyakov-loop component survives in $\langle I \rangle$. 
}
\end{center}
\vspace{-0.5cm}
\end{figure}

Thus, we obtain the analytical relation between 
the Polyakov loop $\langle L_P\rangle$ 
and the Dirac modes in QCD on the temporally odd-number lattice 
\cite{DSI14,SDI16},
\begin{eqnarray}
\langle L_P \rangle=-\frac{(2i)^{N_t-1}}{4N_cV}
\left\langle \sum_n\lambda_n^{N_t-1}\langle n|\hat{U}_4| n \rangle \right\rangle_{\rm gauge~ave.},
\label{eq:main}
\end{eqnarray}
which is mathematically valid 
in both confined and deconfined phases. 
From Eq.(\ref{eq:main}), we can investigate each Dirac-mode contribution 
to the Polyakov loop.
Remarkably, due to the factor $\lambda_n^{N_t -1}$ in Eq.(\ref{eq:main}), 
low-lying Dirac modes gives negligible contribution 
to the Polyakov loop \cite{DSI14,SDI16}.
In lattice QCD simulations, we have numerically confirmed 
the relation (\ref{eq:main}) and scarce contribution of 
low-lying Dirac modes to the Polyakov loop 
in both confined and deconfined phases \cite{DSI14}. 

\section{Polyakov-loop fluctuations and Dirac eigenmodes}

Next, we consider the Polyakov-loop fluctuations, 
which can be a good indicator of the QCD transition \cite{LFKRS13}. 
On the temporally odd lattice, 
we derive Dirac-mode expansion formula for Polyakov-loop fluctuations \cite{DRSS15}, e.g.,
\begin{align}
R_A=
\frac{
\left\langle\left|\sum\lambda_n^{N_t-1} \hat{U}_4^{nn} \right|^2\right\rangle
-
\left\langle\left|\sum\lambda_n^{N_t-1} \hat{U}_4^{nn}\right|\right\rangle^2
}{
\left\langle\left(\sum\lambda_n^{N_t-1}{\rm Re}\left(\mathrm{e}^{2\pi ki/3}\hat{U}_4^{nn}\right)\right)^2\right\rangle
-
\left\langle\sum\lambda_n^{N_t-1}{\rm Re}\left(\mathrm{e}^{2\pi ki/3}\hat{U}_4^{nn}\right)\right\rangle^2
}, 
\end{align}
where $\hat{U}_4^{nn} \equiv \langle n|\hat{U}_4| n \rangle$, and
$k$ is chosen such that the transformed Polyakov loop 
lies in its real sector \cite{DRSS15,LFKRS13}.
The damping factor $\lambda_n^{N_t-1}$ 
appears in the Dirac-mode sum. 
By removing low-lying Dirac modes, the quark condensate rapidly reduces, 
but the Polyakov-loop fluctuation is almost unchanged \cite{DRSS15}.

\section{The Wilson loop and Dirac modes on arbitrary square lattices}

In this section, 
we investigate the string tension and the Dirac modes, 
using the Wilson loop on $R \times T$ rectangle 
on arbitrary square lattices with any number of $N_t$ \cite{SDI16}.
The Wilson loop is expressed by the functional trace, 
\begin{eqnarray}
W \equiv {\rm Tr}_{c} \hat U_1^R \hat U_{-4}^T \hat U_{-1}^R \hat U_4^T
={\rm Tr}_{c} \hat U_{\rm staple} \hat U_4^T, \quad
\hat U_{\rm staple} \equiv \hat U_1^R \hat U_{-4}^T \hat U_{-1}^R.
\end{eqnarray}
For even $T$ (odd $T$ case is similar \cite{SDI16}), 
we consider the functional trace,
\begin{eqnarray}
J \equiv {\rm Tr}_{c,\gamma} \hat U_{\rm staple} \hat {\Slash D}^T
=\sum_{n} \langle n| \hat U_{\rm staple} {\Slash D}^T |n \rangle 
= (-)^{\frac{T}{2}}
\sum_{n} \lambda_n^T \langle n| \hat U_{\rm staple} |n \rangle.~
\end{eqnarray}
Similarly in Sec.~3, one can derive 
$
\langle W \rangle= \frac {(-)^{\frac{T}{2}}2^T}{4}
\left \langle \sum_{n} \lambda_n^T \langle n| \hat U_{\rm staple} |n 
\rangle \right \rangle
$
\cite{SDI16}.
Then, the string tension $\sigma$ is expressed as 
\begin{eqnarray}
\sigma =-\lim_{R,T \to \infty} \frac{1}{RT}{\rm ln} \langle W \rangle
= -\lim_{R,T \to \infty}\frac{1}{RT}
{\rm ln} \left | \left \langle \sum_{n} 
(2 \lambda_n)^T \langle n| \hat U_{\rm staple} |n \rangle \right \rangle \right|.~~
\end{eqnarray}
Because of the factor $\lambda_n^T$ in the sum, 
the string tension $\sigma$ (the confining force) 
is to be unchanged by the removal of 
the low-lying Dirac-mode contribution.

\section{The Polyakov loop and Wilson/clover/domain-wall fermions}

Finally, we express the Polyakov loop with the eigenmodes of 
the Wilson, the clover ($O(a)$-improved Wilson) and the domain-wall fermion kernels, 
where light doublers are absent \cite{R12}.
The clover fermion kernel is given as 
\begin{eqnarray}
\hat K=\frac{1}{2a}\sum_{\mu=1}^{4} \gamma_\mu (\hat U_\mu-\hat U_{-\mu})
+\frac{r}{2a}\sum_{\mu=\pm 1}^{\pm 4} \gamma_\mu (\hat U_\mu-1)
+m+\frac{arg}{2}\sigma_{\mu\nu}G_{\mu\nu},
\label{CFK}
\end{eqnarray}
which becomes the Wilson fermion kernel without the last term in RHS.
We define eigenmodes and eigenvalues of $\hat K$ as 
$\hat K| n \rangle \rangle =i \tilde \lambda_n |n \rangle \rangle$ with 
$\tilde \lambda_n \in {\bf C}$.

We adopt the lattice with $N_t =4l+1$, and consider the functional trace, 
\begin{eqnarray}
J \equiv \Tr(\hat U_4^{2l+1} \hat K^{2l})
=\sum_n\langle \langle n|\hat{U}_4^{2l+1}{\hat{K}}^{2l}|n \rangle \rangle
=\sum_n(i \tilde \lambda_n)^{2l} \langle \langle n|\hat{U}_4^{2l+1}| n \rangle \rangle.
\end{eqnarray}
Since the kernel $\hat K$ in Eq.~(\ref{CFK}) includes many terms, 
$J \equiv \Tr(\hat U_4^{2l+1} \hat K^{2l})$ 
consists of products of link-variable operators. In each product, 
the total number of $\hat U$ does not exceed $N_t$. 
Each product gives a trajectory as Fig.~2.
\begin{figure}[h]
\begin{center}
\includegraphics[scale=0.4]{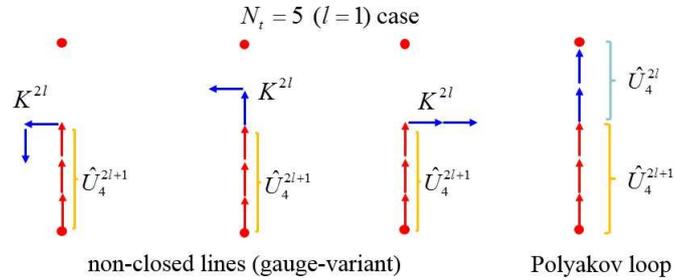}
\caption{
Some examples of the trajectories in 
$J \equiv \Tr(\hat U_4^{2l+1}\hat K^{2l})$. 
The length does not exceed $N_t$. 
Only the Polyakov loop can form a closed loop 
and survives in $\langle J \rangle$. 
}
\end{center}
\vspace{-0.5cm}
\end{figure}
Among the trajectories, however, 
only the Polyakov loop can form a closed loop 
and survives in $\langle J \rangle$, and we derive 
\begin{eqnarray}
\langle L_P \rangle \propto
\langle \sum_n \tilde \lambda_n^{2l}
\langle \langle n|\hat{U}_4^{2l+1}| n \rangle \rangle~\rangle_{\rm gauge~ave.}.
\end{eqnarray}
Due to $\tilde \lambda_n^{2l}$, 
one finds small contribution from low-lying modes of $\hat K$ 
to the Polyakov loop.
We also derive a similar formula for the domain-wall fermion. 

\section{Summary}

We have derived relations between 
the Dirac modes and the confinement quantities
(the Polyakov loop, its fluctuations and the string tension) 
and have found 
scarce contribution from the low-lying Dirac modes. 
This indicates some independence 
of confinement from chiral properties in QCD.

\section*{Acknowledgments}
H.S. and T.M.D. are supported by 
the Grants-in-Aid for Scientific Research 
[Grant No. 15K05076, 15J02108] 
from Japan Society for the Promotion of Science.
The work of K.R. and C.S. is partly
supported by the Polish Science Center 
(NCN) 
under Maestro Grant No. DEC-2013/10/A/ST2/0010.

\end{document}